\documentclass[12pt,a4paper]{article}
\usepackage{physics}
\usepackage{amsfonts}
\usepackage{amsmath}
\usepackage{amssymb}
\usepackage{amsthm}
\usepackage{graphicx}
\usepackage{jheppub}
\usepackage{url}
\usepackage{todonotes}
\usepackage{bm}
\usepackage{xcolor}
\usepackage{caption}
\usepackage{subcaption}
\usepackage{lmodern}
\usepackage[section]{placeins}
\usepackage[utf8]{inputenc}
\usepackage[T1]{fontenc}
\definecolor{refkey}{gray}{0.45}
\definecolor{labelkey}{RGB}{155,48,48}
\hypersetup{
 colorlinks=true,
 citecolor=purple,
 linkcolor=blue,
 urlcolor=violet
 }
 \usepackage[normalem]{ulem}
 
 \newcommand{\mi}{\mathrm{i}}
 \usepackage[bottom]{footmisc}

\newcommand{\bi}{\begin{itemize}}
\newcommand{\ei}{\end{itemize}}

\newcommand{\bea}{\begin{eqnarray}}
\newcommand{\eea}{\end{eqnarray}}
\newcommand{\be}{\begin{equation}}
\newcommand{\ee}{\end{equation}}
\newcommand{\ben}{\begin{eqnarray*}}
\newcommand{\een}{\end{eqnarray*}}
\newcommand{\bem}{\begin{pmatrix}}
\newcommand{\eem}{\end{pmatrix}}
\newcommand{\bl}{\begin{align}}
\newcommand{\el}{\end{align}}
\newcommand{\beg}{\begin{gather}}
\newcommand{\eeg}{\end{gather}}

\newcommand{\cH}{\mathcal{H}}

\newcommand{\IH}{\mathbb{H}}

\renewcommand{\b}{\beta}
\renewcommand{\d}{\delta}

\renewcommand{\r}{\rho}

\newcommand{\TrH[1]}{ {\raise -.5em
           \hbox{$\buildrel {\textstyle {\rm Tr } }\over
{\scriptscriptstyle \cH _ {#1}}$}~}}
\newcommand{\res[1]}{{\raise -.5em 
\hbox{$\buildrel{\textstyle{\rm Res}}\over {\scriptscriptstyle {#1}}$}}}
\newcommand{\tends[1]}{{\raise -.5em 
\hbox{$\buildrel{\longrightarrow}\over {\scriptscriptstyle {#1}}$}}}

\renewcommand{\Im}{\mbox{Im}}
\renewcommand{\Re}{\mbox{Re}}

\def\dbend{\lower3.5pt\hbox{\manual\char127}}

\def\IL{\relax{\rm I\kern-.18em L}}
\def\IH{\relax{\rm I\kern-.18em H}}
\def\rlx{\relax\leavevmode}

\def\ZZ{\rlx\leavevmode\ifmmode\mathchoice{\hbox{\cmss Z\kern-.4em Z}}
 {\hbox{\cmss Z\kern-.4em Z}}{\lower.9pt\hbox{\cmsss Z\kern-.36em Z}}
 {\lower1.2pt\hbox{\cmsss Z\kern-.36em Z}}\else{\cmss Z\kern-.4em
 Z}\fi}

\title{\center \textmd{Quantum Entanglement Index in String Theory}}

\preprint{}
\author[\dagger]{\center Atish Dabholkar,} 
\author[\dagger]{Eleanor Harris,}
\author[*]{and Upamanyu Moitra} 
\emailAdd{atish@ictp.it, eharris@ictp.it, u.moitra@uva.nl}
\affiliation[\dagger]{\begin{center}
Abdus Salam International Centre for Theoretical Physics\\ Strada Costiera 11, Trieste 34151, Italy
\end{center} }
\affiliation[*]{\begin{center}
Institute for Theoretical Physics, Institute of Physics, Universiteit van Amsterdam, Science Park 904, 1089 XH
Amsterdam, The Netherlands
\end{center} }

\abstract{We define a notion of `quantum entanglement index' with the aim to compute it for black hole horizons in string theory at one-loop order using the stringy replica method. We consider the horizon of BTZ black holes to construct the relevant conical orbifolds, labeled by an odd integer $N$, and compute the partition function as a function of $N$, corresponding to the fractional indexed R\'enyi entropy. We show that it is free of tachyons and naturally finite both in the ultraviolet and the infrared, even though it is generically ultraviolet divergent in the field theory limit. Thus, the index provides a useful diagnostic of the entanglement structure of string theory without the need for analytic continuation in $N$.
\vspace{5mm}
}

\keywords{Quantum entanglement, black holes, superstrings}
\makeatletter
\gdef\@fpheader{ \\ }
\makeatother
\begin{document}

\maketitle

\section{Introduction}\label{sec-intro}
In this note, we define a notion of `quantum entanglement index' in quantum field theory and string theory. As we will see, this measure of entanglement is generically ultraviolet--divergent in quantum field theory but is finite both in the ultraviolet (UV) and in the infrared (IR) in string theory. It therefore provides a useful diagnostic of the entanglement structure of string theory as distinct from the Type-III character of quantum field theory. 

To set the stage, we begin with some generalities about entanglement in quantum field theory to motivate this notion. The von Neumann entropy of a partially traced reduced density matrix in a bipartite quantum system is a good measure of the amount of entanglement between the two parts of the system. One may formally regard a quantum field theory in $\mathbb{R}^{1, d} $, with coordinates $(t, x, y^1\ldots , y^{d-1})$, as a bipartite system with the `left' and the `right' parts corresponding to the quantum fields localized to the left ($x < 0$) and the right $(x > 0)$ of a wall at $x = 0$ with $(y^1\ldots , y^{d-1})$ being the coordinates along the wall. We will `compactify' the wall to a $(d-1)$-dimensional torus $\mathbb{T}^{d-1}$ with the identifications $y^i = y^i + L_i$ for $i= 1, \ldots (d-1)$. Given the density matrix $|{0}\rangle \langle{0}|$ of the Minkowski ground state $\ket{0}$, one can obtain the reduced density matrix $\bar\rho_R$ for the right part by the partial trace over the left Hilbert space $\cH_L$, corresponding to the degrees of freedom on the left: 
\begin{equation}
	\bar\r_R := \TrH[_L] |{0}\rangle \langle{0}| \, ,\qquad  \TrH[_R] [\bar\r_R ]=1 \, .
\end{equation}
The unnormalized density matrix $\r_R$ is given by 
\begin{equation}
   \r_R = \exp(-2\pi H_R) \, , \qquad \bar \r_R = \frac{\r_R}{\TrH[_R] [\rho_R]} \, ,
\end{equation}
where $H_R$ is the Rindler Hamiltonian which generates boosts in the `right Rindler wedge' in the $(t, x)$ plane, which is the causal diamond of the right half-line $x > 0$ \cite{Bisognano:1975ih,Bisognano:1976za}. Equivalently, it is the modular Hamiltonian restricted to the right wedge. 

One can now consider a `thermodynamic' partition function with inverse temperature $\b$ and Hamiltonian $H_R$ given by\footnote{The hat distinguishes the spacetime partition function from the worldsheet partition function.}
\begin{equation}\label{partition}
	\hat Z(\b) = \TrH[_R] \left[ e^{-\b H_R} \right] \, , 
\end{equation}
where the trace is over the right Hilbert space $\cH_R$. The von Neumann entropy then equals the `thermodynamic entropy' $S(\b)$ for $\b = 2\pi$, given by
\begin{equation}
	S_{\mathrm{vN}} = \left(- \b \frac{\partial}{\partial \b} + 1 \right) \log \hat Z(\b) \bigg|_{\b =2\pi} = - \TrH[_R]\left[\bar \r_R \log \bar \r_R \right] \, .
\end{equation}
Heuristically, $S_{\mathrm{vN}}$ can be viewed as the entropy of the thermal bath as seen by the Rindler observer in uniform acceleration. In a local quantum field theory, the von Neumann entropy is UV-divergent. 

In supersymmetric quantum field theory, one may wish to generalize these notions to define a `quantum entanglement index' as an analog of the Witten index \cite{Witten:1982df}:
\begin{equation}\label{index}
	\hat Z_{\mathrm{ind}}(\b) = \TrH[_R] \left[ e^{-\b H_R} (-1)^{F_R}\right] = \TrH[_R] \left[ e^{-\b H_R} \, e^{-2\pi \mi J_R}\right]\, ,
	\end{equation}
where $F_R$ is the fermion number in the right Rindler wedge\footnote{We can define the entanglement index for a Rindler wedge because the Rindler modular Hamiltonian is relatively simple. Generalization to more generic spacetime regions may be more involved because modular Hamiltonians in general are more complicated.} and $J_R$ is an $R$-charge which is half-integral for fermions and integral for bosons. In a supersymmetric theory, energy eigenstates with nonzero energy are paired up, leading to a `Bose-Fermi' cancellation, except for the ground state. Hence the indexed partition function is independent of $\b$ and is given by
\begin{equation}
	\hat Z_{\mathrm{ind}}(\b) = d_{\mathrm{ind}} := d_{\rm B} - d_{\rm F} \, ,
\end{equation}
where $d_{\rm B}$ and $d_{\rm F}$ are numbers of bosonic and fermionic ground states respectively. In this note, we will be interested in the situation where there is a unique bosonic ground state, hence $d_{\rm B}=1$ and $d_{\rm F}=0$, but one could consider other generalizations.\footnote{It is well-understood \cite{Sen:2008vm, Dabholkar:2010rm} that ground states associated with the $AdS_2$ horizon of supersymmetric black holes are all bosonic and, correspondingly, $d_{ind}$ is positive and large. We will consider here the horizon of a generic BTZ black hole, not necessarily supersymmetric.} 

All formulae above then generalize easily by replacing $\hat Z$ by $\hat Z_{\mathrm{ind}}$. In particular, the `indexed' von Neumann entropy\footnote{With a suitable choice of $\ket{\phi}$ this may be related to the `pseudo entropy' defined in \cite{Nakata:2020luh}.} is then given by
\begin{equation}\label{indexed-entropy}
	S_{\mathrm{ind}} = \left( - \b \frac{\partial}{\partial \b } +1 \right) \log \hat Z_{\mathrm{ind}}(\b) \bigg|_{\b =2\pi} = \log( d_{\mathrm{ind}} ) \, .
\end{equation}
Note that the indexed entropy need not be positive and may even have an imaginary part. This simply reflects the fact that $d_{\mathrm{ind}}$ may be zero or negative. 

We will be considering string theory in the background of a large Euclidean BTZ black hole at non-zero temperature which breaks supersymmetry because of the boundary conditions around the periodic Euclidean time direction. To correspond to this situation in the quantum field theory setting, it is useful to consider a slight variant by imposing a Scherk-Schwarz boundary condition \cite{Scherk:1978ta}, so that the fermionic fields are antiperiodic and the bosonic fields are periodic around the $y^1$ circle. The energies of the bosonic and fermionic states are then not exactly matched and do not exactly cancel in pairs in the trace. Hence, these boundary conditions slightly break supersymmetry, but in such a way that full supersymmetry is restored in the limit where the $y^1$ circle becomes large. The indexed partition function in this case is expected to have a nontrivial dependence on both $\beta$ and $L_i$. 

The UV divergence arises from the strong quantum correlations between the infinite number of high-energy modes on the left and the right of the wall, close to $x=0$. One may treat the field theory as a quantum system with a finite number of degrees of freedom on a finite lattice. In this case, the lattice spacing $\varepsilon$ provides a suitable ultraviolet cutoff. Equivalently, the divergence can be viewed as that of the thermal entropy of the Rindler heat bath. Each bosonic mode with momentum $k$ and mass $m$ in the thermal bath contributes $-\log \small( 1- e^{-\omega_k/T(x)}\small)$ at one-loop order to $\log \small(\hat Z\small)$ with $\omega^2 = k^2 + m^2$ and where $T(x) = 1/2\pi x$ is the local proper temperature seen by the Rindler observer. Each fermionic mode with momentum $k$, on the other hand, contributes $+\log \small( 1+ e^{-\omega_k/T(x)}\small)$ and hence there is no Bose-Fermi cancellation, even with a supersymmetric spectrum. The local entropy density for a massless field is proportional to $T^d(x)$, and the total entropy is obtained by the integral $A_H \int_\varepsilon^\infty T^{d} (x) dx$, which diverges as $A_H/\varepsilon^{d-1}$ where $A_H = \prod_{i=1}^{d-1} L_i$ is the area of the Rindler horizon. For a massive field, one expects sub-leading corrections of the form $A_H/\varepsilon^{d-1}(m^2 \varepsilon^2 + m^4 \varepsilon^4 + \ldots)$ since the mass can be regarded as small compared to the local temperature to extract the divergence. 

If one instead considers $\log \hat Z_{\mathrm{ind}}$, then each fermion mode would contribute $+\log\small(1 - e^{- \omega_k/T(x)}\small)$ and would cancel against the bosonic contribution if the masses are equal and the spectrum is supersymmetric. Equivalently, at one-loop order, the Bose and Fermi determinants would cancel and this UV-divergence would disappear in a supersymmetric theory.

With Scherk-Schwarz boundary conditions, this cancellation will not be exact. If we dimensionally reduce along the $y^1$ direction, then the bosonic and fermionic fields will have slightly different masses with $\d m^2 \sim 1/L_1^2$. We therefore expect that even if the leading divergence will cancel for the index, the indexed entanglement entropy will be divergent to the sub-leading order with the general form
\begin{equation*}
	S_{\mathrm{ind}} \sim \frac{A_H}{\varepsilon^{d-1}} \left( \d m^2\varepsilon^{2} + \d m^4 \varepsilon^{4} + \ldots \right) \sim \frac{A_H}{\varepsilon^{d-1}}\left(\frac{\varepsilon^{2}}{L_1^2} + \frac{\varepsilon^{4}}{L_1^4} + \ldots \right) \, . 
\end{equation*}
Note that even the partition function $\hat Z_{\rm ind}(\beta)$ by itself would exhibit a similar divergence in a local quantum field theory. It is thus interesting to ask if and how string theory could regulate these divergences. 

We note for completeness that with $\b = 2\pi n$ for $n \in \mathbb{Z}^+$, the partition function \eqref{partition} can be regarded as the R\'enyi partition function of $n$ `replicas' of the original system. It can be evaluated as a path integral over the replicated Euclidean spacetime manifold with a branch cut at $x < 0$. 
The R\'enyi entropy $S(n)$ is then defined by
\begin{equation} \label{Renyi}
S(n) := \frac{1}{1-n}\, \log \TrH[_R][{\bar\r_R}^{\,n}] = \frac{1}{1-n} \left( \log \hat Z(2\pi n) - n \log \hat Z(2\pi) \right) , \qquad n \in \mathbb{Z}^+ \, . 
\end{equation}
With the normalized density matrix with $\Tr (\bar\rho_R) =1$, one expects that $\Tr ({\bar\rho}_R^{\, n})$ is well-defined and analytic in the complex $n$ plane for $\Re(n) \geq 1$. One can seek an analytic continuation to this domain of $S(n)$, defined for positive integers. Under favorable conditions, such an analytic continuation can be unique by Carlson's theorem. The von Neumann entropy is then obtained by 
\begin{equation}
	S_{\mathrm{vN}} = \lim_{n\rightarrow 1} S(n) \, .
\end{equation}

After these preliminary remarks, we now turn to string theory. A method for defining entanglement entropy in perturbative string theory was proposed in \cite{Dabholkar:1994ai,Dabholkar:2001if,Dabholkar:2022mxo} by taking $\b = 2\pi/N$ instead of $2\pi n$, with $N$ an odd positive integer. In this case, the strings live on a conical space with opening angle $2\pi/N$ which can viewed as a $\mathbb{Z}_N$ orbifold of the Euclidean Rindler plane. The worldsheet partition function $Z(N)$ can then be used to define what one might call `fractional R\'enyi entropy' within string theory (order by order in perturbation theory) \cite{Dabholkar:1994ai, Witten:2018xfj}, by identifying 
\begin{equation} \label{Zspacetime}
 Z(N) = \log \hat Z\left(\frac{2\pi}{N}\right) - \log \hat Z(2\pi)= \log \TrH[_R] [\bar\rho_R^{\, 1/N}] \, . 
\end{equation}
Note that in the string theory construction $Z(1)=0$, which is consistent with this identification. This partition function has been computed explicitly for several backgrounds \cite{Dabholkar:2022mxo, Dabholkar:2023ows,Dabholkar:2023yqc, Dabholkar:2023tzd, Dabholkar:2024neq}. These partition functions are naturally finite in the ultraviolet but in all cases suffer from tachyonic divergences in the infrared. In \cite{Dabholkar:2023ows} it was shown that these divergences disappeared upon analytic continuation to the physical region where $0< N \leq 1$. The partition function \eqref{Zspacetime} can thus be used to define a notion of entanglement entropy in string theory that is finite both in the IR and the UV, due to the natural UV regularization of string theory. 

Using the notion of the quantum entanglement index introduced above, we would like to examine the analog of the indexed partition function on the worldsheet. It is more convenient to consider, instead of \eqref{index}, a slightly different partition function
\begin{equation}\label{index2}
	\hat Z_{\rm twist}(\b) = \TrH[_R] \left[ e^{-\b H_R} \, e^{-\b \mi J_R}\right]\, ,
	\end{equation}
with $\beta$ also appearing with the $R$-charge. This modification of the partition function means that the notion of entanglement that we are computing is not the usual one, but can instead be considered as a `twisted' version of it. Similar notions of twisted entanglement entropy in quantum field theory have been considered in different contexts \cite{Nishioka:2013haa, Belin:2013uta, Benini:2024xjv}. One can then proceed to compute the corresponding indexed entropy using a formula analogous to \eqref{indexed-entropy} to obtain
\begin{equation} 
S_{\rm twist} = S_{\rm ind} + \frac{2\pi \mi}{\hat{Z}_{\rm ind} } \langle J_R \rangle = S_{\rm ind}\, .
\end{equation}
The extra term in the first equation vanishes by CPT invariance and hence we could just as well evaluate $\hat Z_{\rm twist}$, which is a simpler quantity. 

We would like to apply the above, rather general, definition of the index to the case of a generic BTZ black hole, which depends on mass and angular momentum.\footnote{The case where supersymmetry is fully preserved corresponds to the zero temperature limit of this, where the contributions of the bosons and fermions exactly cancel. In this fine-tuned case the index will vanish.} We therefore wish to consider $\mathbb{Z}_N$ orbifolds that generalize the construction in \cite{Dabholkar:2023tzd} of horizon orbifolds of Type-IIB string theory on the BTZ black hole, which is locally $AdS_3 \times S^3 \times T^4$. Recall that for the horizon orbifolds in \cite{Dabholkar:2023tzd} the orbifold group acted only in the $AdS_3$ factor, in the plane corresponding to the near horizon Rindler plane, defined by the radial coordinate and the Rindler time coordinate. The spacetime partition function $\hat{Z}(\beta)$ was thus found by taking the $\mathbb{Z}_N$ symmetry as a subgroup of the $U(1)$ Euclidean time-translations in the black hole geometry. The transverse directions $(y^1, \ldots, y^8)$ corresponded to all the remaining directions of the target space. The partition function for the horizon orbifold had tachyonic divergences in the infrared which could be treated by a re-summation and analytic continuation to the physical domain $N \leq 1$, as described in \cite{Dabholkar:2023ows,Dabholkar:2023tzd}. At one-loop order spacetime partition functions are related to worldsheet partition functions by the general relation \eqref{Zspacetime}, whether for the usual partition function or the indexed partition function, and thus we would like to construct the worldsheet partition function corresponding to $\hat{Z}_{\rm twist}(N)$, defined in \eqref{index2}, rather than $\hat{Z}(N)$. For this purpose, one can choose $J_R$ to be one of the $R$-symmetry charges coming from rotations on the $S^3$ or equivalently as a $U(1)$ charge in the $SU(2)$ WZW model. We are thus led to consider a $\mathbb{Z}_N$ action that simultaneously acts on one plane in $AdS_3$ and another plane in the $S^3$ factor. We choose the action such that in the flat limit it corresponds to an action on $\mathbb{R}^4$ belonging to the $SU(2)$ holonomy group, so that one would preserve spacetime supersymmetry as for $\mathbb{C}^2/\mathbb{Z}_N$ asymptotically locally Euclidean (ALE) spaces. The $(y^1, \ldots , y^6)$ directions then correspond to all transverse directions to these two planes. By considering this modification, we find a partition function which has no tachyons, even for $N > 1$. The entanglement index computed in string theory thus provides a measure of entanglement that is naturally finite -- both in the UV, as is natural for a string partition function, but also in the IR due to the lack of tachyonic divergences. The spectrum for superstring theory on such an ($AdS_3 \times S^3)/\mathbb{Z}_N \times T^4$ orbifold was examined in \cite{Martinec:2001cf,Gaberdiel:2023dxt}. Our goal here is to explicitly construct the partition function for computing the fractional indexed R\'enyi entropy. 

In Section \ref{sec-non-susy-summary}, we summarize the results of the partition function for the $\mathbb{Z}_N$ orbifolds of the Euclidean BTZ horizon presented in \cite{Dabholkar:2023tzd}. We then modify this construction in Section \ref{sec-susyBTZ} by taking an appropriate orbifold action that acts simultaneously on the $AdS_3$ and the $S^3$ factors. In Section \ref{sec-Modular}, we demonstrate that one obtains a partition function that is modular invariant as expected. In Section \ref{sec-Flat} we take the flat-space limit to demonstrate that we correctly reproduce expected results on $\mathbb{C}^2/\mathbb{Z}_N \times \mathbb{R}^6$. In this limit, full supersymmetry is restored, and thus the partition function vanishes. Finally, in Section \ref{sec-finiteEE} we show that our partition function, and hence the corresponding indexed fractional R\'enyi entropy, is indeed free of tachyonic IR divergences. It is naturally free of UV divergences. We conclude with a discussion of the implications of our result for the entanglement properties of string theory.

\section{Summary of Horizon Orbifolds \label{sec-non-susy-summary}}
The one-loop superstring partition function for a $\mathbb{Z}_N$ orbifold of the BTZ black hole in $AdS_3 \times S^3 \times T^4$ was obtained in \cite{Dabholkar:2023tzd} by taking the orbifold purely in the $AdS_3$ part of the target space. The result was derived by describing thermal $AdS_3$\footnote{The BTZ black hole can be related to thermal $AdS_3$ by a modular $S$ transformation \cite{Kraus:2006wn}. Here we will stick to the thermal description.} with inverse temperature $\beta$ and chemical potential $\mu$ by an $\mathrm{SL}(2,\mathbb{R})_{\kappa+2}$ WZW model \cite{Gawedzki:1991yu,Maldacena:2000kv}, and it was shown that the orbifold only affects the bosons on the contractible cycle and their fermionic superpartners. We refer to \cite{Dabholkar:2023tzd} for details and simply state the relevant results. The worldsheet partition function is given by 
\begin{equation} \label{WSPF}
 Z(\mathcal{T}, N, \kappa) = \int_\mathcal{D} \frac{d^2 \tau}{\tau_2^2} \, \mathcal{Z}(\tau) \, , 
\end{equation}
where $\mathcal{D}$ is the fundamental domain $|\tau|\geq 1$, $|\tau_1| \leq 1/2$ and $\tau_2>0$, and 
\begin{equation}
\mathcal{Z} (\tau) = \frac{1}{N} \sum_{\substack{m,n\in \mathbb{Z} \\ \ell, k \in \mathbb{Z}_N} } {\cal Z}^{\mathrm{AdS, B}}_{nm; k \ell} (\tau) \qty( {\cal Z}^{\mathrm{F}} {\cal Z}^{\mathrm{ghost}} )_{nm; k\ell} (\tau) {\cal Z}^{S^3} (\tau) {\cal Z}^{T^4} (\tau)\, . \label{pf1}
\end{equation}
The factor of $1/N$ is due to the $\mathbb{Z}_N$ projection of the orbifold with $N$ an odd integer. The sums over twists $k$ and twines $\ell$ in \eqref{pf1} arise from orbifolding, and the $m$ and $n$ sums are related to the momentum and winding modes on the thermal circle. The bosonic contribution to $AdS_3$ is given by
\begin{equation}
{\cal Z}^{\mathrm{AdS, B}}_{nm; k \ell} (\tau)= \frac{\beta \sqrt{\kappa}}{2\pi \ell_s \sqrt{\tau_2}} \frac{\exp( - \frac{(\kappa + 2) \beta^2}{4\pi L^2 \tau_2} |m - n \tau|^2 + \frac{2\pi}{\tau_2} (\Im\ U_{nm; k\ell})^2 )}{|\vartheta_1 ( U_{nm;k\ell} | \tau) |^2} \, , \label{zadsb}
\end{equation}
where $L$ is the $AdS$ radius, $\ell_s$ is the string length, and 
\begin{equation} \label{Unmkl}
 U_{nm, k \ell} = (m - n \tau) \mathcal{T} + \frac{2}{N} (k\tau - \ell) \, . 
\end{equation}
The Jacobi theta function $\vartheta_1 (z | \tau)$ is defined in eq. \eqref{Jacobitheta}. Here, $\mathcal{T}$ is the modular parameter that corresponds to the boundary torus, while the modular parameter $\tau = \tau_1 + \mi \tau_2$ corresponds to the worldsheet torus in the bulk. In particular, $\mathcal{T}$ is related to the inverse temperature and chemical potential by 
\begin{equation}
 \mathcal{T} = \frac{\mathrm{i} \beta}{2\pi L} ( 1 + \mathrm{i} L \mu ) \, . 
\end{equation}
The contribution from the the fermions and ghosts is given by
\begin{equation}
\qty( {\cal Z}^{\mathrm{F}} {\cal Z}^{\mathrm{ghost}} )_{nm; k\ell}= \frac{\tau_2 e^{- \frac{2\pi}{\tau_2} (\mathrm{Im} \, U_{nm; k \ell})^2 }}{|\eta(\tau)|^{4} } \qty| \vartheta_1 \pqty{ \frac{U_{nm; k\ell}}{2} \Big| \tau} |^8 
\end{equation}
for fermions including the $S^3$ and $T^4$ factors. The bosonic contributions to the $S^3$ and $T^4$ partition functions are
\begin{equation} \label{S3orig}
{\cal Z}^{S^3} (\tau) = \sum_{\lambda = 0}^{\kappa-2} | \chi_\lambda^{(\kappa-2)} (\tau, 0 ) |^2 \, , 
\end{equation}
and
\begin{equation}
{\cal Z}^{T^4} (\tau) = \frac{1}{|\eta(\tau)|^8} \sum_{(p_R ,p_L) \in \Gamma_{4,4}} q^{p_R^2/2} \bar{q}^{p_L^2/2} \, , \label{zt4}
\end{equation}
respectively, where $\chi_\lambda^{(\kappa-2)} $ are the $\hat{\mathfrak{su}} (2)_{\kappa-2}$ characters, $q= e^{2\pi \mathrm{i} \tau}$ and $\Gamma_{4,4}$ is the Narain lattice \cite{Narain:1985jj,Narain:1986am}. In this case, neither ${\cal Z}^{S^3} (\tau) $ nor ${\cal Z}^{T^4} (\tau) $ depended on the sums $k\ell$ or $mn$ and could therefore be factored out. Note that in \eqref{zadsb} we have performed the level shift $\kappa \rightarrow \kappa +2$ and similarly in \eqref{S3orig} the shift $\kappa \rightarrow \kappa -2$ which are both necessary to obtain the relevant result for supersymmetry \cite{Dabholkar:2023tzd}. 

\section{Twisted Horizon Orbifolds} \label{sec-susyBTZ}
The one-loop superstring partition function is given by 
\begin{equation} \label{WStrace}
\mathcal{Z}(\tau) = \frac{1}{N} \sum_{k, \ell} \Tr_{\mathcal{H}_k} g^\ell q^{L_0 - \frac{c}{24}} \bar{q}^{\bar{L}_0 - \frac{\bar c}{24} } \, ,
\end{equation}
where $L_0$ and $\bar{L}_0$ are the left and right-moving Virasoro generators, and $\mathcal{H}_k$ is the twisted Hilbert space. Earlier, for the horizon orbifold described in Section \ref{sec-non-susy-summary}, $g$ was an element of the rotation group on the $ AdS_3$ factor alone,
\begin{equation}\label{generator1}
 g = \exp( \frac{4\pi \mathrm{i}}{N} J^{ \mathrm{SL} (2, \mathbb{R} ) } ) \, .
\end{equation}
We recall the reasoning that relates this worldsheet orbifold to the spacetime trace \eqref{partition}. The $\mathbb{Z}_N$ group generated by \eqref{generator1} can be equivalently regarded as the group generated by
\begin{equation}
  g = (-1)^F \exp( \frac{2\pi \mathrm{i}}{N} J^{ \mathrm{SL} (2, \mathbb{R} ) } ) \, , 
\end{equation}
where $F$ is the spacetime fermion number. The Euclidean time evolution is a rotation in the Rindler-$AdS$ plane by $2\pi/N$ and the additional $(-1)^F$ ensures that the fermionic fields are antiperiodic as expected for a spacetime trace. 

To preserve supersymmetry in the flat-space limit, we now want to take $g$ to have a combined action, including on the $S^3$ factor:
\begin{equation} \label{gelements}
 g = \exp( \frac{4\pi \mathrm{i}}{N} J^{ \mathrm{SL} (2, \mathbb{R} ) } ) \otimes \exp( - \frac{4\pi \mi}{N} J^{ \mathrm{SU} (2) } ) \, .
\end{equation}
Because of the additional rotation in the $SU(2)$ factor, now some Green-Schwarz fermions are invariant under the combined action corresponding to the unbroken supersymmetries in the flat limit. On the worldsheet, the Green-Schwarz fermions corresponding to these unbroken supersymmetries will remain periodic. Thus, this new twisted orbifold will correspond to the desired index in spacetime. 

In the twisted orbifold, the partition function \eqref{pf1} is modified in two ways. First, for each twist and twine the three-sphere bosonic partition function \eqref{S3orig} will now be modified as $\mathcal{Z}^{S^3} (\tau) \to \mathcal{Z}^{S^3}_{k\ell} (\tau)$. We discuss the exact form of this piece of the partition function in Section \ref{subsec-S3}. Second, there will be a corresponding modification of the fermionic partition function. For a given twist and twine the bosonic $AdS_3$ piece \eqref{zadsb} will not change. The torus partition function \eqref{zt4} will also remain unchanged. The total partition function of the twisted orbifold is then obtained by summing over twists and twines. Demonstration of the modular invariance of the partition function is straightforward but will have to be suitably modified.

\subsection*{\boldmath Bosons on $\mathbb{Z}_N$ Orbifolds of $S^3$} \label{subsec-S3}
We will now proceed to modify the three-sphere partition function \eqref{S3orig} to include the orbifold twists and twines. The three-sphere partition function is given by the diagonal sum over characters 
\begin{equation} 
 \mathcal{Z}^{S^3}_{k\ell} (\tau, z ) = e^{ - 4 \pi \tau_2 (\kappa -2) \frac{k^2}{N^2} }\sum_{\lambda = 0}^{\kappa - 2} \left| \chi_\lambda^{(\kappa - 2 )} \qty(\tau, z ) \right|^2 \, , \label{zs3kl-inc}
\end{equation}
where 
\begin{equation}
\chi_\lambda^{(\kappa - 2 )} (\tau, z) \equiv \Tr_{\mathcal{H}^{(\kappa - 2)}_{ \lambda} } q^{L_0 - \frac{c}{24} } \, e^{- 2 \pi \mathrm{i} z \, J^{ \mathrm{SU} (2) } } \, . 
\end{equation}
Here $z$ is the chemical potential associated to the $J^{ \mathrm{SU} (2) }$ current that generates the rotation. The exponential prefactor in \eqref{zs3kl-inc} is a consequence of the chiral anomaly\footnote{Equivalently, the prefactor can be considered to arise from a shift in the ground state energy due to the `spectral flow' which takes $L_n \rightarrow L_n + \omega J_n^3 + \frac{\kappa -2}{4}\omega^2$, with $\omega = 2k/N$ \cite{Maldacena:2000hw, Son:2004at, Gaberdiel:2023dxt}. } \cite{Maldacena:2000hw} and is also essential for the modular invariance of the three-sphere partition function, as we shall see in Section \ref{sec-Modular}. The affine $\hat{\mathfrak{su}} (2)_{\kappa-2}$ characters are given by the Weyl-Ka\v{c} character formula \cite{DiFrancesco:1997nk},
\begin{equation} \label{su2char}
\chi_\lambda^{(\kappa - 2 )} (\tau, z) = \frac{\Theta_{\lambda+1, \kappa} (\tau, z)- \Theta_{-\lambda-1, \kappa} (\tau, z) }{\Theta_{1, 2} (\tau, z)- \Theta_{-\-1, 2} (\tau, z) } \, , 
\end{equation}
where the generalized theta functions are
\begin{equation}
\Theta_{\lambda, \kappa} (\tau, z) = \sum_{p \in \mathbb{Z} + \frac{\lambda}{2\kappa} } q^{\kappa p^2} e^{2 \pi \mathrm{i} p \kappa z} \, \label{genth}.
\end{equation}
We would like to take a $\mathbb{Z}_N$ orbifold of the $S^3$ in such a way that supersymmetry is preserved. From the form of \eqref{Unmkl}, we see that we must therefore take $z = U_{00;k\ell} = 2(k \tau - \ell)/N$ in \eqref{zs3kl-inc}. 

\subsection*{Fermionic and Ghost Partition Function} \label{subsec-fermghost}
In the orbifold considered in \cite{Dabholkar:2023tzd}, only one of the five complex fermions was charged under the orbifold action, with boundary conditions
\begin{equation}
\begin{split}
 \psi^1 ( \sigma + 1 , \tau ) &= - e^{2\pi \mathrm{i} a} e^{4\pi \mathrm{i} \frac{k}{N} }\psi^1 ( \sigma , \tau ) \, , \\
 \psi^1 ( \sigma , \tau +1) &= - e^{2\pi \mathrm{i} b} e^{- 4\pi \mathrm{i} \frac{\ell}{N} }\psi^1 ( \sigma , \tau ) \, .
\end{split}
\end{equation}
The holonomies $U_{nm;00}= (m - n \tau)\mathcal{T}$, associated to the $\mathrm{SL}(2,\mathbb{R})$ group action are, on the other hand, treated as a background gauge field, and so appear in the argument of the theta function. Thus, the holomorphic contribution to the partition function from the fermions and the superconformal ghosts was the appropriate GSO projection \cite{Gliozzi:1976qd} of terms 
\begin{equation} \label{fermghostorig}
\begin{split}
 \mathcal{Z}^F \mathcal{Z}^{\text{ghost}} \left[ a \atop b \right] = \frac{\sqrt{\tau_2} } {\eta(\tau)^2} \, \vartheta 
  \left[ a + \frac{2k}{N} \atop b - \frac{2\ell}{N} \right] (U_{nm;00} | \tau) \, \vartheta^3 \left[ a \atop b \right] (0 | \tau) \, . 
 \end{split}
\end{equation}

In the case at hand, one additional complex fermion $\psi^2$, stemming from the three-sphere, will be charged under the orbifold action only, with the three other fermions remaining uncharged; $\psi^i ( \sigma + 1 , \tau ) = - e^{2\pi \mathrm{i} a} \psi^i ( \sigma , \tau )$, $\psi^i ( \sigma , \tau +1) = - e^{2\pi \mathrm{i} b} \psi^i ( \sigma , \tau )$ for $i = 3,4,5$. Thus, \eqref{fermghostorig} gets modified to 
\begin{equation} 
\begin{split}
 \mathcal{Z}^F \mathcal{Z}^{\mathrm{ghost}} \left[ a \atop b \right] = \frac{\sqrt{\tau_2} } {\eta(\tau)^2} \, \vartheta 
  \left[ a + \frac{2k}{N} \atop b - \frac{2l}{N} \right] (U_{nm;00} | \tau) \, \vartheta \left[ a + \frac{2k}{N}\atop b - \frac{2\ell}{N} \right] (0 | \tau) \, \vartheta^2 \left[ a \atop b \right] (0 | \tau) \, . 
 \end{split}
\end{equation}
The contribution to the partition function for a given winding mode, momentum, twist and twine sector is then given by taking the GSO projection, which is implemented by the sum 
\begin{equation} \label{GSOferm}
\begin{split}
 \qty( {\cal Z}^{\mathrm{F}} {\cal Z}^{\mathrm{ghost}} )_{nm; k\ell} &= \frac{e^{- \frac{2\pi}{\tau_2} (\mathrm{Im} \, U_{nm;00})^2 }}{4} \left|\sum_{a,b = 0}^{1/2} s_{ab}(k,l) \, \mathcal{Z}^F \mathcal{Z}^{\mathrm{ghost}} \left[ a \atop b \right] \right|^2 \, , 
\end{split}
\end{equation}
where the spin structure constants, $s_{ab}(k,l) = (-1)^{2a + 2b + 4ab } e^{ - 8\pi \mathrm{i} k b /N}$, are chosen to ensure modular invariance \cite{Blumenhagen:2013fgp}, and the exponential prefactor in \eqref{GSOferm} is due to the chiral anomaly in the path integral measure \cite{Alvarez-Gaume:1986rcs}. Using the identity 
\begin{equation} 
 \vartheta \left[ a + x\atop b + y\right](z|\tau) = e^{\pi \mathrm{i} \tau x^2 +2\pi \mathrm{i} x(y + z + b) }
\vartheta \left[ a \atop b \right] (x \tau + y +z | \tau )
  \, , 
\end{equation}
as well as the Riemann theta identity (see e.g. \cite{Dabholkar:1994ai, Dabholkar:2023yqc}), we can then perform the sum in \eqref{GSOferm} to arrive at 
\begin{equation}
 \qty( {\cal Z}^{\mathrm{F}} {\cal Z}^{\mathrm{ghost}} )_{nm; k\ell} = \frac{\tau_2 e^{ - \frac{2\pi}{\tau_2} (\mathrm{Im} \, U_{nm; k \ell})^2 - 8 \pi \tau_2 \frac{k^2}{N^2} }}{|\eta(\tau)|^{4} } \qty| \vartheta_1^2 \pqty{ \tfrac{U_{nm; 00}}{2} \Big| \tau} \vartheta_1^2 \pqty{ \tfrac{U_{nm; (2k) (2\ell) }}{2} \Big| \tau} |^2 \, ,
\end{equation}
where $\vartheta_1(z|\tau)$ is the Jacobi theta function, defined in \eqref{Jacobitheta}. 

\subsection*{Superstring Partition Function}
In summary, combining all the various contributions, we have that the partition function for the supersymmetric orbifold is 
\begin{equation} \label{stringPFfinal}
\mathcal{Z} (\tau) = \frac{1}{N} \sum_{\substack{m,n\in \mathbb{Z} \\ \ell, k \in \mathbb{Z}_N} } {\cal Z}^{\mathrm{AdS, B}}_{nm; k \ell} (\tau) \qty( {\cal Z}^{\mathrm{F}} {\cal Z}^{\mathrm{ghost}} )_{nm; k\ell} (\tau) {\cal Z}^{S^3}_{k\ell} (\tau) {\cal Z}^{T^4} (\tau) \, , 
\end{equation}
with
\begin{equation} \label{AbsBosonPF}
{\cal Z}^{\mathrm{AdS, B}}_{nm; k \ell} (\tau)= \frac{\beta \sqrt{\kappa}}{2\pi \ell_s \sqrt{\tau_2}} \frac{\exp( - \frac{(\kappa + 2) \beta^2}{4\pi L^2 \tau_2} |m - n \tau|^2 + \frac{2\pi}{\tau_2} (\Im\ U_{nm; k\ell})^2 )}{|\vartheta_1 ( U_{nm;k\ell} | \tau) |^2} \, , 
\end{equation}
\begin{equation} \label{Fgh2}
( {\cal Z}^{\mathrm{F}} {\cal Z}^{\mathrm{ghost}} )_{nm; k\ell} (\tau) = \frac{\tau_2 e^{ - \frac{2\pi}{\tau_2} (\mathrm{Im} \, U_{nm; k \ell})^2 - 8 \pi \tau_2 \frac{k^2}{N^2} }}{|\eta(\tau)|^{4} } \qty| \vartheta_1^2 \pqty{ \tfrac{U_{nm; 00}}{2} \Big| \tau} \vartheta_1^2 \pqty{ \tfrac{U_{nm; (2k) (2\ell) }}{2} \Big| \tau} |^2 \, , 
\end{equation}
 \begin{equation} \label{S3PF2}
 \mathcal{Z}^{S^3}_{k\ell} (\tau) = e^{ - 4 \pi \tau_2 (\kappa - 2) \frac{k^2}{N^2} } \sum_{\lambda = 0}^{\kappa-2} \left| \chi_\lambda^{(\kappa-2)} \qty(\tau, \frac{2(k \tau - \ell)}{N} ) \right|^2 \, , 
\end{equation}
and
\begin{equation}
{\cal Z}^{T^4} (\tau) = \frac{1}{|\eta(\tau)|^8} \sum_{(p_R ,p_L) \in \Gamma_{4,4}} q^{p_R^2/2} \bar{q}^{p_L^2/2} \, . 
\end{equation}

\section{Modular Invariance \label{sec-Modular}}
In this section, we check the modular invariance of the superstring partition function. We start by examining the modular properties of the three-sphere and the fermion-ghost pieces in a given twist and twine sector.

Under modular transformations, $U_{nm, k \ell} = (m - n \tau) \mathcal{T} + \frac{2}{N} (k\tau - \ell)$ transforms as 
\begin{equation}
  \begin{split}
    S: \quad U_{nm, k \ell} &\rightarrow -\frac{1}{\tau} U_{m(-n);\ell (-k)} \, ,\\
    T: \quad U_{nm, k \ell} &\rightarrow U_{n(m-n);k(\ell-k)} \, , 
  \end{split}
\end{equation}
where $S$ takes $\tau \rightarrow -1/\tau$ and $T$ shifts $\tau \rightarrow \tau +1$. 

Modular invariance of the bosonic $AdS_3$ piece of the partition function was demonstrated in \cite{Dabholkar:2023tzd}. Similarly here, we must consider modular transformations of the bosonic three-sphere partition function for a given twist and twine sector, \eqref{S3PF2}. The modular transformations of the $\hat{\mathfrak{su}} (2)_{\kappa}$ characters are
\begin{equation}
\begin{aligned} \label{ZAdSB2}
\chi^{(\kappa)}_\lambda \qty(- \frac{1}{\tau} , \frac{z}{\tau}) &= \exp \left(\frac{\pi \mathrm{i} \kappa z^2}{2\tau} \right) \sum_{\lambda' = 0}^\kappa S_{\lambda \lambda'} \chi^{(\kappa)}_{\lambda'} (\tau, z) \, , \\
\chi^{(\kappa)}_\lambda (\tau+1, z) &= \exp(2 \pi \mathrm{i} \frac{2 \lambda (\lambda +2) - \kappa }{8 (\kappa +2)} ) \chi^{(\kappa)}_\lambda (\tau, z) \, ,\\
\end{aligned}
\end{equation}
where $S_{\lambda \lambda'}$ is the standard modular $S$-matrix,
\begin{equation}
S_{\lambda \lambda'} = \sqrt{ \frac{2}{\kappa +2} } \sin( \frac{\pi (\lambda + 1)(\lambda' +1) }{\kappa +2} ) \, .
\end{equation}
Therefore, under the modular $S$-transformation we have 
\begin{equation}
\begin{aligned}
{\cal Z}^{S^3}_{k\ell} \qty( -\frac{1}{\tau} ) &=  e^{ - 4 \pi \frac{\tau_2}{|\tau|^2} (\kappa -2)\frac{k^2}{N^2} } \sum_{\lambda = 0}^{\kappa - 2} \left|  \chi_\lambda^{(\kappa - 2)} \qty( -\tfrac{1}{\tau} , \tfrac{ 2(-k - \ell \tau)}{N \tau } ) \right|^2 \\
&= e^{ - 4 \pi \frac{\tau_2}{|\tau|^2} (\kappa -2)\frac{k^2}{N^2} } \sum_{\lambda = 0}^{\kappa - 2} \left| e^{2 \pi \mi (\kappa -2) \frac{(\ell \tau+k )^2}{N^2 \tau}  } \sum_{\lambda'} S_{\lambda \lambda'} \chi_{\lambda'}^{(\kappa - 2)} \qty(\tau , \tfrac{ 2(-\ell \tau - k)}{N} ) \right|^2 \\ 
&=  \mathcal{Z}^{S^3}_{\ell (-k)} (\tau) \, .\\
\end{aligned}
\end{equation}
In deriving the above we have made use of the unitarity of the modular $S$-matrix and the fact that the characters are even under $z\rightarrow -z$. Since the $T$-transformation only shifts each character by a phase, it is even simpler to see that 
\begin{equation}
{\cal Z}^{S^3}_{k\ell} (\tau+1) =  \mathcal{Z}^{S^3}_{k(\ell-k)} (\tau)  \, .
\end{equation}
Similarly, one can show that the modular properties of the fermion and ghost piece of the partition function for a given twist and twine sector are 
\begin{equation}
\begin{split}
  \qty( {\cal Z}^{\mathrm{F}}  {\cal Z}^{\mathrm{ghost}} )_{nm; k\ell} \qty( - \frac{1}{\tau} ) &= \qty( {\cal Z}^{\mathrm{F}}  {\cal Z}^{\mathrm{ghost}} )_{m(-n);\ell (-k)} (\tau) \, , \\
  \qty( {\cal Z}^{\mathrm{F}}  {\cal Z}^{\mathrm{ghost}} )_{nm; k\ell} (\tau + 1) &= \qty( {\cal Z}^{\mathrm{F}}  {\cal Z}^{\mathrm{ghost}} )_{n(m-n); k(\ell - k)} (\tau) \, . 
\end{split}
\end{equation}
Using these results, it is straightforward to show that the full superstring partition function \eqref{stringPFfinal} is modular invariant:
\begin{equation} 
\begin{split}
\mathcal{Z} \qty( - \frac{1}{\tau} ) &= \frac{1}{N} \sum_{\substack{m,n\in \mathbb{Z} \\ \ell, k \in \mathbb{Z}_N} } {\cal Z}^{\mathrm{AdS, B}}_{m(-n);\ell (-k)} (\tau) \qty( {\cal Z}^{\mathrm{F}}  {\cal Z}^{\mathrm{ghost}} )_{m(-n);\ell (-k)} (\tau) {\cal Z}^{S^3}_{\ell(-k)} (\tau) {\cal Z}^{T^4} (\tau) \, , 
\end{split}
\end{equation}
which is exactly $\mathcal{Z} \qty( \tau )$ since we can use the $\mathbb{Z}_N$ symmetry to shift $k \rightarrow - k$. The invariance under the $T$-transformation follows similarly. 
\section{The Flat-Space Limit \label{sec-Flat}}
In this section, we calculate the flat-space limit of our final answer for the one-loop partition function with fixed $\beta/L$. Since supersymmetry is left unbroken in our current set-up, we expect that the answer will vanish due to the fermionic zero modes. However, the different components of the answer are worth examining, since we also expect that in the flat limit the contributions from the bosons and fermions will exactly cancel. This will provide a check of our result. The common curvature radius of both $AdS_3$ and $S^3$ is related to the WZW level by $L = \sqrt{\kappa} \ell_s$, where $\ell_s$ is the string scale. We are therefore interested in the $\kappa \rightarrow \infty$ limit. In this limit, the target space becomes
\begin{equation}
(\mathrm{AdS}_3 \times S^3) / \mathbb{Z}_N \to \mathbb{C}^2 / \mathbb{Z}_N \times \mathbb{R}^2.
\end{equation}
As in \cite{Dabholkar:2023tzd}, we first perform the unfolding trick \cite{Polchinski:1985zf} to trade the sum over $n$ for an integral over the strip:
\begin{equation}
\int\limits_{\mathcal{D}} \sum_{m, n \in \mathbb{Z} } =\int\limits_{\mathcal{S}} \sum_{\substack{m \in \mathbb{Z}_{<0 } \\ n = 0 }} + \int\limits_{\substack{\mathcal{D} \\ m = 0 = n} }\, , \label{unfold}
\end{equation}
where $\mathcal{D}$ is the fundamental domain defined by $\tau_2 >0$, $|\tau_1| \leq 1/2$ and $|\tau| \geq 1$, and $\mathcal{S}$ is the strip with $\tau_2 > 0$ and $|\tau_1 | \leq 1/2$, containing infinite copies of the fundamental domain. From the form of the exponential in \eqref{AbsBosonPF}, we can see that the $m\neq 0$ terms on the right-hand side of \eqref{unfold} will be suppressed in the $\kappa \to \infty$ limit. The $S^3$ partition function does not depend on the winding or momenta, but it does depend on $\kappa$, so we will have to examine its flat-space limit separately.

The $m = 0 = n$ term of the first two components of the partition function is 
\begin{equation}
\begin {aligned}
{\cal Z}^{\mathrm{AdS, B}}_{00; k \ell} ( {\cal Z}^{\mathrm{F}}  {\cal Z}^{\mathrm{ghost}} )_{00; k\ell} (\tau) &= \frac{\beta \sqrt{\kappa \tau_2}}{2\pi\ell_s} e^{- 8 \pi \tau_2 \frac{k^2}{N^2}  } \left| \frac{ \vartheta_1^2 \pqty{ \frac{U_{00; 00}}{2} \Big| \tau} \vartheta_1^2 \pqty{ \frac{U_{00; (2k) (2\ell) }}{2}  \Big| \tau} }{ \eta^2 (\tau) \vartheta_1 ( U_{00;k\ell} | \tau) } \right|^2 \\
&= \frac{\beta \sqrt{\kappa \tau_2}}{2\pi\ell_s} e^{- 8 \pi \tau_2 \frac{k^2}{N^2}  } \left| \frac{ \vartheta_1^2 \pqty{ 0 | \tau} \vartheta_1^2 \pqty{ \frac{2(k\tau - \ell)}{N} \Big| \tau} }{ \eta^2 (\tau) \vartheta_1 \pqty{ \frac{2(k\tau - \ell)}{N} \Big| \tau} } \right|^2 \, . \label{fll00k}
\end{aligned}
\end{equation}
We already see the fermionic zero modes in the $\vartheta_1 (0| \tau)$ terms in the numerator. The large volume limit of the torus partition function is 
\begin{equation}
  {\cal Z}^{T^4} (\tau) \rightarrow \frac{V_4 }{|\eta(\tau)|^8 \ell_s^4 \tau_2^2}\, , 
\end{equation}
where we have normalized the volume of the torus to be $(2\pi)^4 V_4$. We now come to the $S^3$ part of the partition function. It will be useful to rewrite the expression \eqref{S3PF2} as 
\begin{equation}
 \mathcal{Z}^{S^3}_{k\ell} (\tau) = \frac{e^{ - 4 \pi \tau_2 (\kappa - 2) \frac{k^2}{N^2} } }{ \left| \vartheta_1 \pqty{ z| \tau} \right|^2 } \sum_{\lambda = 0}^{\kappa-2} \left| \sum_{ p \in \mathbb{Z} + \frac{\lambda+1}{2\kappa} } q^{ \kappa p^2 } ( e^{2\pi \mi p \kappa z} - e^{-2\pi \mi p \kappa z} ) \right|^2 \, ,  \label{zs3two}
\end{equation}
where $z = 2(k \tau - \ell)/N$, and we have rewritten the denominator using
\begin{equation}
\Theta_{1, 2} (\tau, z)- \Theta_{-\-1, 2} (\tau, z) = - \mi \vartheta_1 ( z | \tau) \, . 
\end{equation}
The term corresponding to the limit $k , \ell \rightarrow 0$ must be treated carefully. The expression for this case was previously shown in \cite{Dabholkar:2023tzd} to be
\begin{equation}
 \mathcal{Z}^{S^3}_{00} (\tau) = \frac{1}{4\pi | \eta( \tau) |^6}  \pqty{ \frac{\kappa}{\tau_2} }^{\frac{3}{2}} \, . 
\end{equation}
For the twisted sectors we shift the inner sum of \eqref{zs3two} to get 
\begin{equation}
  \sum_{p \in \mathbb{Z}} e^{ - 2 \pi \mi \kappa \tau \frac{k^2}{N^2} }\bqty{ e^{2\pi \mi \kappa \tau \pqty{ p + \frac{\lambda+1}{2\kappa} + \frac{k}{N}  }^2 - \frac{2\pi \mi \ell (2p \kappa + \lambda +1 ) }{N} } - e^{2\pi \mi \kappa \tau \pqty{ p + \frac{\lambda+1}{2\kappa} - \frac{k}{N}  }^2 + \frac{2\pi \mi \ell (2p \kappa + \lambda+1) }{N} } } \, .
  \label{psumch}
\end{equation}
As $\kappa \to \infty$, the sum will collapse to a single term for which the coefficient of $2\pi i \tau$ in the exponent, either $(p + \frac{\lambda+1}{2\kappa} + \frac{k}{N} )^2$ or $(p + \frac{\lambda+1}{2\kappa} - \frac{k}{N} )^2$, is a minimum. The $\mathbb{Z}_N$ symmetry allows us to choose the range 
\begin{equation}
  -(N-1)/2 \leq k \leq (N-1)/2 \, . 
\end{equation}
In this range of values the $p = 0$ term dominates the sum. For $k>0$, the second term is the leading contribution and for $k< 0$ the first term is the leading contribution. For $k=0$ both terms contribute evenly. We can then write $\lambda +1 =\sqrt{\kappa} x$, and approximate the sum over $\lambda$ by an integral. In the $k=0$ case, this becomes 
\begin{equation}
\begin{split}
  \mathcal{Z}^{S^3}_{0\ell} (\tau) &\approx \frac{4 \sqrt{\kappa} }{ \left| \vartheta_1 \pqty{ -\frac{2\ell}{N}\big| \tau} \right|^2 } \int_0^\infty e^{ - \pi \tau_2 x^2 } \sin^2 \pqty{ 2\pi x \sqrt{\kappa} \frac{\ell}{N} } \\
  &= \frac{1 }{ \left| \vartheta_1 \pqty{ -\frac{2\ell}{N}\big| \tau} \right|^2 } \sqrt{\frac{\kappa}{\tau_2}} \pqty{ 1 - \exp( - \frac{4\pi \kappa}{\tau_2} \frac{\ell^2}{N^2} ) } \, . \label{kzero}
\end{split}
\end{equation}
For non-zero $\ell$, the second term disappears under the limit $\kappa \to \infty$. For $k\neq 0$, the sum \eqref{psumch} also collapses to the single term $p=0$, which we can write uniformly as 
\begin{equation}
\pm \, e^{-2\pi \mi \kappa \tau \frac{k^2}{N^2} } e^{ 2\pi \mi \kappa \tau \pqty{ \frac{\lambda +1}{2\kappa} - \frac{|k|}{N} }^2 + \mathrm{sgn}(k) \frac{2\pi \mi \ell (\lambda+1)}{N}  } \, . 
\end{equation}
Since we take the modulus-squared of this quantity, the last term in the exponent does not contribute, and so we can similarly approximate this as the integral 
\begin{equation}
\begin{split}
  \mathcal{Z}^{S^3}_{k\ell} (\tau) &\approx \frac{e^{ 8 \pi \tau_2 \frac{k^2}{N^2} } \sqrt{\kappa}}{ \left| \vartheta_1 \pqty{ \frac{ 2(k \tau - \ell)}{N} \big| \tau} \right|^2 } \int_0^\infty \dd{x} \exp[ - 4\pi \tau_2 \pqty{ \frac{x}{2} - \frac{\sqrt{\kappa}|k|}{N} }^2 ] \\
  &= \frac{1}{2} \frac{e^{ 8 \pi \tau_2 \frac{k^2}{N^2} } }{ \left| \vartheta_1 \pqty{ \frac{ 2(k \tau - \ell)}{N} \big| \tau} \right|^2 }  \sqrt{ \frac{\kappa}{\tau_2} } \bqty{1 + \mathrm{Erf} \qty(2 \sqrt{\pi \tau_2 \kappa} \frac{|k|}{N} ) } \, .
\end{split}
\end{equation}
In the limit $\kappa \to \infty$, the error function takes the value $1$ and therefore, we obtain
\begin{equation}
 \mathcal{Z}^{S^3}_{k\ell} (\tau) = \frac{ e^{8\pi \tau_2 \frac{k^2}{N^2} } }{ \left| \vartheta_1 \pqty{ \frac{ 2(k \tau - \ell)}{N} \Big| \tau} \right|^2 } \sqrt{ \frac{\kappa}{\tau_2} } \, ,
\end{equation}
which clearly agrees with \eqref{kzero} for $k = 0$. As expected, this result matches the flat-space partition function for a bosonic field. The exponential term involving $\exp(4\pi \kappa \tau_2 k^2 / N^2)$ has canceled out. The exponential term left behind precisely cancels that from the fermionic term in \eqref{fll00k}. We thus obtain the flat-space limit of the full partition function,
\begin{equation}
\mathcal{Z} (\tau) = \frac{1}{N} \sum_{ \ell, k \in \mathbb{Z}_N} \frac{\beta \kappa V_4 }{2\pi \tau_2^2 \ell_s^{5}} \left| \frac{ \vartheta_1^2 \pqty{ 0 | \tau} \vartheta_1^2 \pqty{ \frac{2(k\tau - \ell)}{N} \Big| \tau} }{ \eta^6 (\tau) \vartheta_1^2 \pqty{ \frac{2(k\tau - \ell)}{N} \Big| \tau} } \right|^2 .
\end{equation}
As one would expect, the fermionic and bosonic orbifold contributions cancel one another in the Green-Schwarz formalism.

\section{Finite entanglement entropy}
\label{sec-finiteEE}

It is easy to see that the partition function of the orbifolds we have constructed is free from the tachyonic divergences that are present in the non-supersymmetric case. We describe it in some detail below.

Note that for the contribution coming from the three-sphere, apart from in the denominator, $\tau$ and $\kappa$ mostly appear together in the partition function \eqref{zs3two}. The large $\tau_2$ analysis of this part of the partition function will therefore follow similarly to the flat-space limit. By again using the unfolding trick described in \eqref{unfold} we can focus on the terms in the sum with $n=0$. Leaving aside the $T^4$ contribution for the moment, the relevant bosonic and fermionic terms are therefore approximately 
\begin{equation} \label{unfoldedPF}
  ({\cal Z}^{\mathrm{AdS, B}} {\cal Z}^{\mathrm{F}}  {\cal Z}^{\mathrm{ghost}} \mathcal{Z}^{S^3})_{0m; k \ell} \approx
  \frac{\beta \kappa e^{ - \frac{(\kappa + 2) \beta^2 m^2 }{4\pi L^2 \tau_2} }}{2\pi \ell_s |\eta(\tau)|^{4} } \left| \frac{\vartheta_1^2 \pqty{ \tfrac{U_{0m; 00}}{2} \Big| \tau} \vartheta_1^2 \pqty{ \tfrac{U_{0m; (2k) (2\ell) }}{2}  \Big| \tau} }{\vartheta_1 \pqty{ U_{00;k\ell} | \tau} \vartheta_1 ( U_{0m;k\ell} | \tau) } \right|^2 \, ,
\end{equation}
where, as a reminder, $U_{nm, k \ell} = (m - n \tau) \mathcal{T} + \frac{2}{N} (k\tau - \ell)$. We can then see from the explicit form of the Jacobi theta function
\begin{equation} \label{Jacobitheta}
  \vartheta_1(z|\tau) = \sum_{j \in \mathbb{Z}} q^{\frac{1}{2}\left(j + \frac{1}{2}\right)^2} e^{2\pi \mi \left( j + \frac{1}{2}\right)\left( z +\frac{1}{2} \right)} \, , 
\end{equation}
that in the $\tau_2\rightarrow \infty$ limit, the sum will collapse to the $j=0$ term, and therefore the theta functions will not contribute any tachyonic divergences, since 
\begin{equation} \label{Finite}
  \frac{\qty| \vartheta_1^2 \pqty{ \tfrac{U_{0m; 00}}{2} \Big| \tau} \vartheta_1^2 \pqty{ \tfrac{U_{0m; (2k) (2\ell) }}{2}  \Big| \tau} |^2}{\left|\vartheta_1 ( U_{0m;k\ell} | \tau) \vartheta_1 \pqty{ U_{00;k\ell} | \tau} \right|^2} \approx \left| \frac{e^{- 4\pi \tau_2 k/N }}{e^{- 4\pi \tau_2 k/N}} \right|^2 \, .
\end{equation}
We therefore indeed have a partition function that is free from IR divergences. Being a string theory partition function, it is also naturally free of UV divergences. 

We conclude that the fractional indexed R\'enyi entropy is finite in string theory\footnote{The spectrum of $A{\rm dS}_3$ is known to contain ``long string'' states that arise from additional representations that are generated by spectral flow \cite{Maldacena:2000hw,Maldacena:2000kv}. These spectrally flowed states are related to poles of the partition function in the middle of moduli space; see \cite{Dabholkar:2023tzd}. The divergences stemming from them are related to the infinite $AdS_3$ volume and are not relevant to the current discussion.} even though it is divergent in quantum field theory. This clearly indicates that the entanglement structure of string theory is very different from the Type-III character exhibited by quantum field theory. Note also that the factional R\'enyi entropy with $N> 1$ corresponds to the unphysical region $0<n<1$. Since it is finite in this region one would typically expect even better behavior, and hence finiteness, in the region $n>1$, if one is able to find an analytic continuation. 

Even though we have analyzed the quantum entanglement index \eqref{index} only at one-loop level, in certain contexts the result may in fact be one-loop exact. Therefore, since it is possibly topological, the index could also be relevant for probing strongly coupled dynamics and dualities, as is the case with the Witten index and its generalizations such as the elliptic genus. Moreover, while the index is divergent in quantum field theory, it may still provide useful new information for explorations of entanglement entropy as is the case with the central charge in 
$(1+1)$-dimensional conformal field theory and its holographic duals. 

\acknowledgments

We would like to thank Valentin Benedetti and Sameer Murthy for useful discussions. UM thanks the ICTP for its warm hospitality during the progress of this work. UM is supported by the European Research Council under the European
Union’s Seventh Framework Programme (FP7/2007-2013), ERC Grant agreement
ADG 834878.

\appendix

\bibliographystyle{JHEP}
\bibliography{entangle.bib}

\end{document}